# Resource Allocation: Realizing Mean-Variability-Fairness Tradeoffs

Vinay Joseph, Gustavo de Veciana and Ari Arapostathis
The University of Texas at Austin
vinayjoseph@mail.utexas.edu, gustavo@ece.utexas.edu, ari@ece.utexas.edu

*Abstract*—Network Utility Maximization (NUM) provides a key conceptual framework to study reward allocation amongst a collection of users/entities in disciplines as diverse as economics, law and engineering. In network engineering, this framework has been particularly insightful towards understanding how Internet protocols allocate bandwidth, and motivated diverse research efforts on distributed mechanisms to maximize network utility while incorporating new relevant constraints, on energy, power, storage, stability, etc., e.g., for systems ranging from communication networks to the smart-grid. However when the available resources and/or users' utilities vary over time, reward allocations will tend to vary, which in turn may have a detrimental impact on the users' overall satisfaction or quality of experience.

This paper introduces a generalization of the NUM framework which incorporates the detrimental impact of temporal variability in a user's allocated rewards. It explicitly incorporates tradeoffs amongst the mean and variability in users' reward allocations, as well as fairness across users. We propose a simple online algorithm to realize these tradeoffs, which, under stationary ergodic assumptions, is shown to be asymptotically optimal, i.e., achieves a long term performance equal to that of an offline algorithm with knowledge of the future variability in the system. This substantially extends work on NUM to an interesting class of relevant problems where users/entities are sensitive to temporal variability in their service or allocated rewards.

## I. Introduction

Network Utility Maximization (NUM) provides a key conceptual framework to study (fair) reward allocation among a collection of users/entities across disciplines as diverse as economics, law and engineering. For example, [25] introduces NUM in the context of realizing fair allocations of a *fixed* amount of water $c$ to $N$ farms. The amount of water $w_i$ allocated to the $i$th farm is a resource which yields a reward $r_i = f_i(w_i)$ to the $i$th farm. Here, $f_i$ is a concave function mapping allocated water (resource) to yield (reward), and these can differ across farms. The allocation maximizing $\sum_{1 \leq i \leq N} r_i$ is a reward (utility) maximizing solution to the problem. Fairness can be imposed on the allocation by changing the objective of the problem to $\sum_{1 \leq i \leq N} U(r_i)$ for an appropriately chosen concave function $U$. Now, suppose that we have to make the allocation decisions periodically to respond to time varying water availability $(c_t)_{t \in \mathbb{N}}$ and utility functions $(f_{i,t})_t$. Then, subject to the time varying constraints, one could

This research was supported in part by Intel and Cisco under the VAWN program, and by the NSF under Grant CNS-0917067.

maximize (see for e.g., [29], [16])

$$\sum_{1 \leq i \leq N} U(\overline{r}_i) \qquad (1)$$

to obtain a resource allocation scheme which is fair in the delivery of time average reward $\overline{\mathbf{r}}$.

In network engineering, the NUM framework has served as a particularly insightful setting to study (reverse engineer) how the Internet's congestion control protocols allocate bandwidth, how to devise schedulers for wireless systems with time varying channel capacities, and also motivated the development of distributed mechanisms to maximize network utility in diverse settings including communication networks and the smart grid, while incorporating new relevant constraints, on energy, power, storage, power control, stability, etc.

When the available resources/rewards and/or users' utilities vary over time, reward allocations amongst users will tend to vary, which in turn may have a detrimental impact on the users' utility or perceived service quality. In fact, temporal variability in farm water availability can even have a negative impact on crop yield (see [27]). This motivates modifications of formulations with objectives such as the one in (1) to account for this impact.

Indeed temporal variability in utility, service, rewards or associated prices are particularly problematic when humans are the eventual recipients of the allocations. Humans typically view temporal variability negatively, as a sign of an unreliable service, network or market instability, or as a service, which when viewed through human's cognitive and behavioral responses, has a degraded Quality of Experience (QoE). This in turn can lead users to make decisions, e.g., change provider, act upon perceived market instabilities, etc., which can have serious implications on businesses and engineered systems, or economic markets. For problems involving resource allocation in networks, [5] argues that predictable or consistent QoS is essential and even points out that it may be appropriate to intentionally lower the quality delivered to the user if that level is sustainable.

For a user viewing a video stream, variations in video quality over time have a detrimental impact on the user's QoE, see e.g., [31], [15], [23]. Indeed [31] suggested that variations in quality can result in a QoE that is worse than that of a constant quality video with lower average quality. Furthermore, [31] proposed a metric for QoE given below which penalizes the temporal standard deviation of the quality:

Mean Quality $- \kappa \sqrt{\text{Temporal Variance in Quality}}$

2where $\kappa$ is an appropriately chosen positive constant. [9] and [30] argue that less variability in the service processes can improve customer satisfaction by studying data for large retail banks and major airlines respectively. Aversion towards temporal variability is not just restricted to human behavior, for instance, see [22] for a discussion of the impact of temporal variability in nectar reward on foraging behavior of bees. Also, variability in resource allocation in networks can lead to burstiness which can degrade network performance (see [7], [24]). These examples illustrate the need for extending the NUM framework to incorporate the impact of variability.

This paper introduces a generalized NUM framework which explicitly incorporates the detrimental impact of temporal variability in a user's allocated rewards. We use the term rewards as a proxy for the resulting utility of, or any other quantity associated with, allocations to users/entities in a system. Our goal is to explicitly tackle the task of incorporating tradeoffs amongst the mean and variability in users' rewards. Thus, for example, in a variance-sensitive NUM setting, it may make sense to reduce a user's mean reward so as to reduce his/her variability. As will be discussed in the sequel, there are many ways in which temporal variations can be accounted for, and which, in fact, present distinct technical challenges. In this paper, we shall take a simple elegant approach to the problem which serves to address systems where tradeoffs amongst the mean and variability over time need to be made rather than systems where the desired mean (or target) is known (as in minimum variance control, see [2]), or where the issue at hand is minimization of the variance of a cumulative reward at the end of a given (e.g., investment) period.

To better describe the characteristics of the problem we introduce some preliminary notation. We shall consider a network shared by a set $\mathcal{N}$ of users (or other entities) where $N := |\mathcal{N}|$ denotes the number of users in the system. Throughout the paper, we distinguish between random variables (and random functions) and their realizations by using upper case letters for the former and lower case for the latter. We use bold letters to denote vectors, e.g., $\mathbf{a} = (a_i)_{i \in \mathcal{N}}$. We let $(b)_{1:T}$ denote the finite length sequence $(b(t))_{1 \leq t \leq T}$ (in the space associated with the objects of the sequence). Let $\mathbb{N}$ denote set of natural numbers $\{1, 2, 3, ...\}$. Let $\mathbb{N}$, $\mathbb{R}$ and $\mathbb{R}_+$ denote the sets of positive integers, real numbers and nonnegative real numbers respectively. For any function $U$ on $\mathbb{R}$, let $U'$ denote its derivative.

Let $r_i(t)$ represents the reward allocated to user $i$ at time $t$. Then $\mathbf{r}(t) = (r_i(t))_{i \in \mathcal{N}}$ is the vector of rewards to users $\mathcal{N}$ at time $t$, and $(\mathbf{r})_{1:T}$ represents the rewards allocated over time slots $t = 1, \ldots, T$ to the same users. We assume that reward allocations are subject to time varying network constraints,

$$c_t(\mathbf{r}(t)) \leq 0 \text{ for } t = 1, \ldots, T,$$

where each $c_t : \mathbb{R}^N \to \mathbb{R}$ is a convex function, thus implicitly defining a convex set of feasible reward allocations. To formally capture the impact of the time-varying rewards on users' QoE consider the following *offline* convex optimization problem OPT(T):

$$\max_{(\mathbf{r})_{1:T}} \sum_{i \in \mathcal{N}} U_i^E \left( \overbrace{\underbrace{m^T(r_i)}_{\text{Mean Reward}} \underbrace{-U_i^V\left(\text{Var}^T(r_i)\right)}_{\text{Penalty for Variability}}}^{\text{User } i\text{'s QoE}} \right),$$

subject to $\quad c_t(\mathbf{r}(t)) \leq 0, \ \mathbf{r}(t) \geq \mathbf{0} \ \forall \, t \in \{1, ..., T\},$

where the functions $m^T(\cdot)$, $\text{Var}^T(\cdot)$ and $e_i^T(\cdot)$ are defined next.

**Definition 1.** For $(a(t))_{t \in \mathbb{N}} \in \mathbb{R}^{\mathbb{N}}$, let

$$m^T(a) := \frac{1}{T} \sum_{t=1}^{T} a(t),$$

$$\text{Var}^T(a) := \frac{1}{T} \sum_{t=1}^{T} \left( a(t) - m^T(a) \right)^2,$$

$$e_i^T(a) := m^T(a) - U_i^V\left(\text{Var}^T(a)\right).$$

We will also (abusing notation) use the above operators on any finite sequence $(a)_{1:T} \in \mathbb{R}^T$.

We refer to OPT(T) as an offline optimization because time-varying time constraints $(c_t)_{1:T}$ are assumed to be known. Here, we allow increasing functions $\left(U_i^E, U_i^V\right)_{i \in \mathcal{N}}$ that ensure that the above optimization problem is convex. For user $i$, the argument of the function $U_i^E$ is our proxy for the user's QoE. Thus, the desired fairness in the allocation of QoE across the users can be imposed by appropriately choosing $\left(U_i^E\right)_{i \in \mathcal{N}}$. Note that the first term $m^T(r_i)$ in user $i$'s QoE is the user's mean reward allocation, whereas the presence of the empirical variance function $\text{Var}^T(r_i)$ in the second term penalizes temporal variability in a reward allocation. Further, flexibility in picking $\left(U_i^V\right)_{i \in \mathcal{N}}$ allows for several different ways to penalize such variability. Indeed, one can in principle have a variability penalty that is convex or concave in variance. Hence, the formulation OPT($T$) allows us to realize tradeoffs among mean, fairness and variability associated with the reward allocation by appropriately choosing the functions $\left(U_i^E, U_i^V\right)_{i \in \mathcal{N}}$.

*A. Main contributions*

The main contribution of this paper is the development of a *simple asymptotically optimal online* algorithm, Adaptive Variability-aware Reward allocation (AVR), for realizing mean-variance-fairness tradeoffs. The algorithm requires almost no statistical information about the system, and its characteristics are as follows:

(i) in each time slot, $c_t$ is revealed, and AVR allocates rewards by solving optimization problem OPT-ONLINE given below:

$$\max_{\mathbf{r}} \sum_{i \in \mathcal{N}} \left(U_i^E\right)'(e_i(t)) \left( r_i - \left(U_i^V\right)'(v_i(t))(r_i - m_i(t))^2 \right)$$

subject to $c_t(\mathbf{r}) \leq 0, \ \mathbf{r} \geq 0,$

where $e_i(t) = m_i(t) - U_i^V(v_i(t))$ for each $i \in \mathcal{N}$ is an estimate of the user's QoE based on estimated means and variances $\mathbf{m}(t)$ and $\mathbf{v}(t)$; and,

(ii) it updates (vector) parameters $\mathbf{m}(t)$ and $\boldsymbol{v}(t)$ to keep track of the mean and variance respectively associated with the reward allocation under AVR.

Under stationary ergodic assumptions on the time-varying constraints $(C_t)_{t\in\mathbb{N}}$, we show that our *online* algorithm AVR is asymptotically optimal, i.e., achieves a performance equal to that of the *offline* optimization OPT(T) introduced earlier as $T \to \infty$. This is a strong optimality result, which at first sight may be surprising due to the rewards on $\text{Var}^T(\cdot)$ and the time varying nature of the constraints $(c_t)_{t\in\mathbb{N}}$. The key idea is to exploit the characteristics of the problem, by keeping online estimates for the relevant quantities associated with users' allocations, e.g., the mean and variance which over time are shown to converge, and this eventually enables the online policy to produce allocations corresponding to the optimal stationary policy. Proving this result is somewhat challenging as it requires showing that the estimates based on allocations produced by our online policy, AVR, (which itself depends on the estimated quantities), will converge to the desired values. To our knowledge this is the first attempt to generalize the NUM framework in this direction. We contrast our problem formulation and approach to past work in the literature addressing 'variability' minimization, risk-sensitive control and other MDP based frameworks in the related work below.

### B. Related Work

Network Utility Maximization (NUM) provides the key conceptual framework to study how to fairly allocate rewards amongst a collection of users/entities. The work in [25] provides a network-centric overview of NUM. All the work on NUM including several major extensions (for e.g., [14], [29], [28], [21] etc.) have ignored the impact of variability in reward allocation. Our work [12] is one of the first to tackle network resource allocation incorporating the impact of variability explicitly. In particular, we addressed a special case of the problem studied in this paper that only allows for linear functions $\left(U_i^E, U_i^V\right)_{i\in\mathcal{N}}$, and an asymptotically optimal online resource allocation algorithm for a wireless network supporting video streaming users is proposed. The algorithm proposed and analyzed in this paper is a generalization of gradient based algorithms studied in [1], [16] and [29]. However, our approach for proving asymptotic optimality of such simple online gradient based schemes for 'convex' resource allocation problems (with objectives involving certain types of time averages) is an important generalization of the approaches in [29] and [13]. In [29], the focus is on objectives such as (1), and does not allow for the addition of terms like temporal variance to the objective. The approach in [13] relies on the use of results on sensitivity analysis of optimization problems, and only allows for linear $\left(U_i^E\right)_{i\in\mathcal{N}}$ and concave $\left(U_i^V\right)_{i\in\mathcal{N}}$.

Adding a temporal variance term in the cost takes the objective out of the basic dynamic programming setting (even when $\left(U_i^E, U_i^V\right)_{i\in\mathcal{N}}$ are all linear) as the overall cost is not decomposable over time, i.e., can not be written as a sum of costs each depending only on the allocation at that time- this essentially is what makes sensitivity to variability challenging. For risk sensitive decision making, MDP based approaches aimed at realizing optimal tradeoffs between mean and temporal variance in reward/cost were proposed in [8] and [26]. While they consider a more general setting than ours where actions can even affect the process $(C_t)_{t\in\mathbb{N}}$, the approaches proposed in these works suffer from the curse of dimensionality as they require solving large optimization problems. For instance, the approach in [8] involves solving a quadratic program in the (typically large) space of state-action frequencies. Note that these approaches for risk sensitive decision making are different from ones focusing on the variance of the *cumulative* cost/reward such as the one in [19].

Variability or perceived variability could be measured in many different ways, and temporal variance considered in this paper is one of them. One could also 'reduce variability' using a minimum variance controller (see [2]) where we have certain target reward values fixed ahead of time and big fluctuations from these targets are undesirable. Note however that in using this approach, we have to fix our targets ahead of time, and thus lose the ability to realize tradeoffs between the mean and variability in reward allocation. One could also measure variability using switching costs like in [18], which considers the problem of achieving tradeoffs between average cost and time average switching cost associated with data center operation, and proposes algorithms with good performance guarantees for adversarial scenarios. The decision regarding how to measure variability should ultimately be based on the application setting under consideration.

### C. Organization of the paper

Section II introduces the system model and assumptions. In Section III, we present and study the offline formulation for optimal variance sensitive joint reward allocation OPT(T). We start Section IV by formally introducing our online algorithm AVR and present a convergence result associated with it. This in turn serves as the basis for establishing the asymptotic optimality of AVR. Section V is devoted to the proof of AVR's convergence. We conclude the paper in Section VI. Proofs for some of the results presented in these sections are discussed in the appendices.

## II. SYSTEM MODEL

We consider a slotted system where time slots are indexed by $t \in \mathbb{N}$, and the system serves a fixed set of users $\mathcal{N}$ and let $N := |\mathcal{N}|$.

We assume that rewards are allocated subject to time varying constraints. The reward allocation $\mathbf{r}(t) \in \mathbb{R}_+^N$ in time slot $t$ is constrained to satisfy the following inequality

$$c_t(\mathbf{r}(t)) \leq 0,$$

where $c_t$ denotes the realization of a randomly selected function $C_t$ from a (arbitrarily large) finite set $\mathcal{C}$ of real valued maps on $\mathbb{R}_+^N$. We make the following assumptions on these constraints:

**Assumptions C1-C3 (Time varying constraints on rewards)**

**C.1** $(C_t)_{t\in\mathbb{N}}$ is a stationary ergodic process.

**C.2** Feasible region corresponding to each constraint is bounded: there is a constant $0 < r_{\max} < \infty$ such that for any $c \in \mathcal{C}$ and $\mathbf{r} \in \mathbb{R}_+^N$ satisfying $c(\mathbf{r}) \leq 0$, we have $r_i \leq r_{\max}$ for each $i \in \mathcal{N}$. [1]

**C.3** Each function $c \in \mathcal{C}$ is convex and differentiable on an open set containing $[0, r_{\max}]^N$ with $c(\mathbf{0}) \leq 0$ and

$$\min_{\mathbf{r} \in [0, r_{\max}]^N} c(\mathbf{r}) < 0. \tag{2}$$

Let $(\pi(c))_{c \in \mathcal{C}}$ denote the marginal distribution associated with the stationary ergodic process $(C_t)_{t \in \mathbb{N}}$, and let $C^\pi$ denote a random constraint with distribution $(\pi(c))_{c \in \mathcal{C}}$.

As pointed out in C.1, we model the evolution of the constraints over time as a stationary ergodic process. We view $(C_t)_{t \in \mathbb{N}}$ as a random process where each $C_t$ can be interchangeably viewed as a random function or an index selected randomly from a finite set $\mathcal{C}$. If condition C.2 holds, then we can upper bound any feasible allocation under any constraint in $\mathcal{C}$ using $r_{\max} \mathbf{1}_N$ where $\mathbf{1}_N$ is the $N$ length vector with each component equal to one. Condition C.3 ensures that the feasible sets are convex, and the differentiability requirement simplifies the exposition. The remaining requirements in C.3 are useful in studying the optimization problem OPT($T$).

Next we discuss the assumptions on the functions $(U_i^V)_{i \in \mathcal{N}}$ associated with the variability penalties. Let $v_{\max} := r_{\max}^2$.

**Assumptions U.V: (Variability penalty)**

**U.V.1**: For each $i \in \mathcal{N}$, $U_i^V$ is well defined and differentiable on an open set containing $[0, v_{\max}]$ satisfying $\min_{v \in [0, v_{\max}]} (U_i^V)'(v) > 0$, and $(U_i^V)'(\cdot)$ is Lipschitz continuous.

**U.V.2**: For each $i \in \mathcal{N}$ and any $z_1, z_2 \in [-\sqrt{v_{\max}}, \sqrt{v_{\max}}]$ with $z_1 \neq z_2$, and $\alpha \in (0, 1)$ with $\bar{\alpha} = 1 - \alpha$, we have

$$U_i^V\left((\alpha z_1 + \bar{\alpha} z_2)^2\right) < \alpha U_i^V(z_1^2) + \bar{\alpha} U_i^V(z_2^2). \tag{3}$$

Note that any non-decreasing (not necessarily strictly) convex function satisfies (3), but the condition is weaker than a convexity requirement. For instance, using triangle inequality, one can that $U_i^V(v_i) = \sqrt{v_i + \delta}$ for $\delta > 0$ satisfies all the conditions described above for any $v_{\max}$[2]. This function is not convex but is useful as it transforms variance to (approximately) the standard deviation for small enough $\delta > 0$. We will later see that our algorithm (Section I-A) can be simplified if any of the functions $U_i^V$ are linear. Hence, we define the following subsets of $\mathcal{N}$:

$$\mathcal{N}_l := \{i \in \mathcal{N} : U_i^V \text{ is linear}\},$$
$$\mathcal{N}_n := \{i \in \mathcal{N} : U_i^V \text{ is not linear}\}.$$

Next we discuss the assumptions on the functions $(U_i^E)_{i \in \mathcal{N}}$ used to impose fairness associated with the QoE across users. Recall that the proxy for QoE for user $i$ is $e_i(t) = m_i(t) - U_i^V(v_i(t))$ and, let

$$e_{\min,i} := -U_i^V(v_{\max}) \text{ and } e_{\max,i} := r_{\max} - U_i^V(0).$$

---

[1] We could allow the constant $r_{\max}$ to be user dependent. But, we avoid this for notational simplicity.

[2] We need such approximations because $U_i^V(v_i) = \sqrt{v_i}$ violates U.V.1

**Assumption U.E: (Fairness in QoE)**

**U.E**: For each $i \in \mathcal{N}$, $U_i^E$ is concave and differentiable on an open set containing $[e_{\min,i}, e_{\max,i}]$ with $(U_i^E)'(e_{\max,i}) > 0$, and $(U_i^E)'(\cdot)$ is Lipschitz continuous.

Note that concavity and the condition $(U_i^E)'(e_{\max,i}) > 0$ ensure that $(U_i^E)'$ is strictly positive on $[e_{\min,i}, e_{\max,i}]$. For each $i \in \mathcal{N}$, although $U_i^E$ has to be defined over an open set containing $[e_{\min,i}, e_{\max,i}]$, only the definition of the function over $\left[-U_i^V(0), e_{\max,i}\right]$ affects the optimization. This is because we can achieve this value of QoE for each user just by allocating $0\mathbf{1}_N$ in each time slot. Thus, for example, we can choose any function from the following class of strictly concave increasing functions parametrized by $\alpha \in (0, \infty)$ ([20])

$$U_\alpha(e) = \begin{cases} \log(e) & \text{if } \alpha = 1, \\ (1-\alpha)^{-1} e^{1-\alpha} & \text{otherwise}, \end{cases} \tag{4}$$

and can satisfy U.E by making minor modifications to the function. For instance, we can use the following modification $U^{E,\log}$ of the log function for any (small) $\delta > 0$: $U^{E,\log}(e) = \log(e - e_{\min,i} + \delta)$, $e \in [e_{\min,i}, e_{\max,i}]$. The above class of functions are commonly used to enforce fairness specifically to achieve allocations that are $\alpha-$fair (see [25]). A larger $\alpha$ corresponds to a more fair allocation which eventually becomes max-min fair as $\alpha$ goes to infinity.

*Applicability of the model*

We close this section by illustrating the wide scope of the framework discussed above by describing examples of scenarios that fit it nicely. The presence of time-varying constraints $c_t(\mathbf{r}) \leq 0$ allows us to apply the model to several interesting and useful settings. In particular, we discuss three wireless network settings and show that the model can handle problems involving time-varying exogenous loads and time-varying utility functions.

*1) Time varying capacity constraints:* We start by discussing the case where the reward in a time slot is the rate allocated to the user in that time slot. Let $\mathcal{P}$ denote a finite (but arbitrarily large) set of positive vectors where each vector corresponds to the peak transmission rates achievable to the set of users in a given time slot. Let $\mathcal{C} = \left\{c_\mathbf{p} : c_\mathbf{p}(\mathbf{r}) = \sum_{i \in \mathcal{N}} \frac{r_i}{p_i} - 1, \mathbf{p} \in \mathcal{P}\right\}$. Here, for any allocation $\mathbf{r}$, $r_i/p_i$ is the fraction of time the wireless system needs to serve user $i$ in time slot $t$ to deliver data at the rate of $r_i$ when the user has peak transmission rate $p_i$. Thus, the constraint $c_\mathbf{p}(\mathbf{r}) \leq 0$ can be seen as a scheduling constraint that corresponds to the requirement that the sum of the fractions of time that different users are served in a time slot should be less than or equal to one. We can verify that we satisfy C.2-C.3 by choosing $r_{\max} = \max_{\{\mathbf{p} \in \mathcal{P}, i \in \mathcal{N}\}} p_i$ and noting that $c(\delta_{\text{feas}} \mathbf{1}_N) < 0$ for $\delta_{\text{feas}} = \frac{1}{2N} \min_{\{\mathbf{p} \in \mathcal{P}, i \in \mathcal{N}\}} p_i$.

*2) Time-varying exogenous constraints:* We can also allow for time varying exogenous constraints on the wireless system by appropriately defining the set $\mathcal{C}$. For instance, suppose a base station in a cellular network allocates rates to users some of whom are streaming videos. As pointed above, the QoE of users viewing video content is sensitive to temporal variability in quality. But, while allocating rates to these users, we may also have to account for the time varying resource requirements of the voice and data traffic handled by the base station. We can model this by defining

$$\mathcal{C} = \left\{ c_{\mathbf{p},f} : c_{\mathbf{p},f}(\mathbf{r}) = \sum_{i \in \mathcal{N}} \frac{r_i}{p_i} - (1-f), \ \mathbf{p} \in \mathcal{P}, f \in \mathcal{T}_{fr} \right\}$$

where $\mathcal{T}_{fr}$ is a finite set of real numbers in $[0,1)$ where each element in the set corresponds to a fraction of a time slot's time that is allocated to other traffic. Let $f_{\max} = \max_{f \in \mathcal{T}_{fr}} f$. Then, we can verify that we satisfy C.2-C.3 by choosing $r_{\max} = \max_{\{\mathbf{p} \in \mathcal{P}, i \in \mathcal{N}\}} p_i$ and noting that $c(\delta_{\text{feas}} \mathbf{1}_N) < 0$ for $\delta_{\text{feas}} = \frac{1}{2N} \min_{\{\mathbf{p} \in \mathcal{P}, i \in \mathcal{N}\}} p_i (1 - f_{\max})$.

*3) Time varying utility functions:* For users streaming video content, it is more appropriate to view the perceived video quality of a user in a time slot as the reward for that user in that slot. However, for users streaming video content, the dependence of perceived video quality [3] on the compression rate is time varying. This is typically due to the possibly changing nature of the content, e.g., from an action to a slower scene. Hence, the 'utility' function that maps the reward (i.e., perceived video quality) derived from the allocated reward (i.e., the rate) is time varying. This setting can be handled as follows. Let $q_{t,i}(\cdot)$ denote the strictly increasing concave function that, in time slot $t$, maps the rate allocated to user $i$ to user perceived video quality. For each user $i$, let $\mathcal{Q}_i$ be a finite set of such functions. This setting can be modeled by set of constraints:

$$\mathcal{C} = \left\{ c_{\mathbf{p},\mathbf{q}} : c_{\mathbf{p},\mathbf{q}}(\mathbf{r}) = \sum_{i \in \mathcal{N}} \frac{q_i^{-1}(r_i)}{p_i} - 1, \right.$$
$$\left. \mathbf{p} \in \mathcal{P}, q_i \in \mathcal{Q}_i \ \forall \ i \in \mathcal{N} \right\}.$$

Note that each element in $\mathcal{C}$ is a convex function. If we assume that each function $q \in \mathcal{Q}$ is differentiable and convex, then we can verify that we satisfy C.2-C.3 by choosing $r_{\max} = \max_{\{\mathbf{p} \in \mathcal{P}, i \in \mathcal{N}, q \in \mathcal{Q}\}} q(p_i)$ and a small enough $\delta_{\text{feas}}$ so that $c(\delta_{\text{feas}} \mathbf{1}_N) < 0$.

To summarize, our framework allows substantial freedom in modeling temporal variability in both the available resources and the sensitivity of the users' reward/utility to their allocations, as well as fairness across users' QoE.

## III. OPTIMAL VARIANCE-SENSITIVE OFFLINE POLICY

In this section, we study OPT(T), the offline formulation for optimal joint reward allocation introduced in Section I. In the offline setting, we assume that $(c)_{1:T}$, the realization of the process $(C)_{1:T}$, is known. We denote the objective function of OPT(T) by $\phi_T$, i.e.,

$$\phi_T(\mathbf{r}) := \sum_{i \in \mathcal{N}} U_i^E \left( e_i^T(r_i) \right),$$

---

[3] in a short duration time slot roughly a second long which corresponds to a collection of 20-30 frames

where $e_i^T(\cdot)$ is as in Definition 1. Hence the optimization problem OPT($T$) can be rewritten as:

$$\max_{(\mathbf{r})_{1:T}} \phi_T(\mathbf{r}) \tag{5}$$
$$\text{subject to} \quad c_t(\mathbf{r}(t)) \leq 0 \ \forall \ t \in \{1, ..., T\}, \tag{6}$$
$$r_i(t) \geq 0 \ \forall \ t \in \{1, ..., T\}, \forall \ i \in \mathcal{N}. \tag{7}$$

The next result asserts that OPT($T$) is a convex optimization problem satisfying Slater's condition (Section 5.2.3, [6]) and that it has a unique solution.

**Lemma 1.** *OPT($T$) is a convex optimization problem satisfying Slater's condition with a unique solution.*

*Proof:* By Assumptions U.E and U.V, the convexity of the objective of OPT($T$) is easy to establish once we prove the convexity of the function $U_i^V(\text{Var}^T(\cdot))$ for each $i \in \mathcal{N}$. Using (3) and the definition of $\text{Var}^T(\cdot)$, we can show that $U_i^V(\text{Var}^T(\cdot))$ is a convex function for each $i \in \mathcal{N}$. The details are given next. For any two quality vectors $(\mathbf{r^1})_{1:T}$ and $(\mathbf{r^2})_{1:T}$, any $i \in \mathcal{N}$, $\alpha \in (0,1)$ and $\bar{\alpha} = 1 - \alpha$, we have that

$$\text{Var}^T\left(\alpha r_i^1 + \bar{\alpha} r_i^2\right)$$
$$= \frac{1}{T} \sum_{t=1}^T \left( \alpha \left( r_i^1(t) - m^T(r_i^1) \right) + \bar{\alpha} \left( r_i^2(t) - m^T(r_i^2) \right) \right)^2$$
$$\leq \left( \sqrt{\frac{1}{T} \sum_{t=1}^T \left( \alpha \left( r_i^1(t) - m^T(r_i^1) \right) \right)^2} \right.$$
$$\left. + \sqrt{\frac{1}{T} \sum_{t=1}^T \left( \bar{\alpha} \left( r_i^2(t) - m^T(r_i^2) \right) \right)^2} \right)^2$$
$$= \left( \alpha \sqrt{\text{Var}^T(r_i^1)} + \bar{\alpha} \sqrt{\text{Var}^T(r_i^2)} \right)^2 \tag{8}$$

where the above inequality follows from triangle inequality for the Euclidean norm. Using this, (3) and the monotonicity of $U_i^V$, we have

$$U_i^V \left( \text{Var}^T \left( \alpha r_i^1 + \bar{\alpha} r_i^2 \right) \right)$$
$$\leq \alpha U_i^V \left( \text{Var}^T(r_i^1) \right) + \bar{\alpha} U_i^V \left( \text{Var}^T(r_i^2) \right). \tag{9}$$

Thus, $U_i^V(\text{Var}^T(\cdot))$ is a convex function. Thus, by the concavity of $U_i^E(\cdot)$ and $-U_i^V(\text{Var}^T(\cdot))$, we can conclude that OPT($T$) is a convex optimization problem. Also, from (8) and (3) (since we have strict inequality), we can conclude that we have equality in (9) only if

$$\text{Var}^T(r_i^1) = \text{Var}^T(r_i^2), \tag{10}$$

or equivalently

$$r_i^1(t) = r_i^2(t) + m^T(r_i^1) - m^T(r_i^2) \ \forall t \in \{1, ..., T\}. \tag{11}$$

Further, Slater's condition is satisfied and it mainly follows from (2) in Assumption C.3.

Now, for any $i \in \mathcal{N}$, $U_i^E$ and $-U_i^V(\text{Var}^T(\cdot))$ are not necessarily strictly concave. But, we can still show that OPT($T$) has a unique solution. Let $(\mathbf{r^1})_{1:T}$ and $(\mathbf{r^2})_{1:T}$ be two optimal solutions to OPT($T$). Then, from the concavity of the objective, $\left(\alpha \left(r_i^1\right)_{1:T} + \bar{\alpha}\left(r_i^2\right)_{1:T}\right)$ is also an optimal solution for any $\alpha \in (0,1)$ and $\bar{\alpha} = 1 - \alpha$. Due to convexity of $U_i^E(\cdot)$ and $U_i^V\left(\text{Var}^T(\cdot)\right)$, this is only possible if for each $i \in \mathcal{N}$ and $1 \leq t \leq T$,

$$U_i^V\left(\text{Var}^T\left(\alpha r_i^1 + \bar{\alpha} r_i^2\right)\right) = \alpha U_i^V\left(\text{Var}^T\left(r_i^1\right)\right) + \bar{\alpha} U_i^V\left(\text{Var}^T\left(r_i^2\right)\right).$$

Hence (11) and (10) hold. Due to optimality of $\left(\mathbf{r}^1\right)_{1:T}$ and $\left(\mathbf{r}^2\right)_{1:T}$, we have that

$$\sum_{i \in \mathcal{N}} U_i^E\left(\frac{1}{T} \sum_{t=1}^T r_i^2(t) - U_i^V\left(\text{Var}^T\left(r_i^2\right)\right)\right)$$
$$= \sum_{i \in \mathcal{N}} U_i^E\left(\frac{1}{T} \sum_{t=1}^T r_i^1(t) - U_i^V\left(\text{Var}^T\left(r_i^2\right)\right)\right)$$
$$= \sum_{i \in \mathcal{N}} U_i^E\left(\frac{1}{T} \sum_{t=1}^T r_i^2(t) + m^T\left(r_i^1\right) - m^T\left(r_i^2\right) - U_i^V\left(\text{Var}^T\left(r_i^2\right)\right)\right)$$

where the first equality follows from (10) and the second one follows from (11). Since $U_i^E$ is a strictly increasing function for each $i \in \mathcal{N}$, the above equation implies that $m^T\left(r_i^1\right) = m^T\left(r_i^2\right)$ and thus (using (11)) $\mathbf{r}^1(t) = \mathbf{r}^2(t)$ for each $t$ such that $1 \leq t \leq T$. From the above discussion, we can conclude that OPT($T$) has a unique solution. ∎

We let $\left(\mathbf{r}^T\right)_{1:T}$ denote the optimal solution to OPT($T$). Since OPT($T$) is a convex optimization problem satisfying Slater's condition (Lemma 1), the Karush-Kuhn-Tucker (KKT) conditions ([6]) given next hold.

**KKT-OPT(T)**:
There exist nonnegative constants $\left(\mu^T\right)_{1:T}$ and $\left(\gamma^T\right)_{1:T}$ such that for all $i \in \mathcal{N}$ and $t \in \{1,...,T\}$, we have

$$\left(U_i^E\right)'\left(e_i^T\left(r_i^T\right)\right)\left(\frac{1}{T} - \frac{2\left(U_i^V\right)'\left(\text{Var}^T\left(r_i^T\right)\right)}{T}\left(r_i^T(t) - m^T\left(r_i^T\right)\right)\right) - \frac{\mu^T(t)}{T} c'_{t,i}(\mathbf{r}^T(t)) + \frac{\gamma_i^T(t)}{T} = 0, \quad (12)$$

$$\mu^T(t) c_t(\mathbf{r}^T(t)) = 0, \quad (13)$$
$$\gamma_i^T(t) r_i^T(t) = 0. \quad (14)$$

Here $c'_{t,i}$ denotes $\frac{\partial c_t}{\partial r_i}$, and we have used the fact that for any $t \in \{1,...,T\}$

$$\frac{\partial}{\partial r(t)}\left(T\text{Var}^T(r)\right) = 2\left(r(t) - m^T(r)\right).$$

From (12), we see that the optimal reward allocation $\mathbf{r}^T(t)$ in any time time slot $t$ depends on the entire allocation $\left(\mathbf{r}^T\right)_{1:T}$ only through the following three quantities associated with $\left(\mathbf{r}^T\right)_{1:T}$: (i) the time average rewards $\mathbf{m}^T$, (ii) $\left(\left(U_i^E\right)'\right)_{i \in \mathcal{N}}$ evaluated at the quality of experience of the respective users, (iii) $\left(\left(U_i^V\right)'\right)_{i \in \mathcal{N}}$ evaluated at the variance seen by the respective users. So, if these time averages associated with the *optimal solution* were somehow known, the optimal allocation for each time slot $t$ could be determined by solving an optimization problem (derived from the KKT conditions) that only requires these time averages, and knowledge of $c_t$ (associated with current time slot) and not entire $(c)_{1:T}$. We exploit this key idea in formulating our online algorithm in the next section.

## IV. ADAPTIVE VARIANCE AWARE REWARD ALLOCATION

The reward allocations for AVR are obtained by solving the following problem denoted OPTAVR($\mathbf{m}, \mathbf{v}, c$):

$$\max_{\mathbf{r}} \sum_{i \in \mathcal{N}} \left(U_i^E\right)'(e_i)\left(r_i - \left(U_i^V\right)'(v_i)(r_i - m_i)^2\right)$$

$$\text{subject to} \quad c(\mathbf{r}) \leq 0, \quad (15)$$
$$r_i \geq 0 \quad \forall\, i \in \mathcal{N}, \quad (16)$$

where $e_i = m_i - U_i^V(v_i)$ for each $i \in \mathcal{N}$. Here $\mathbf{m}$, $\mathbf{v}$ and $\mathbf{e}$ correspond to current estimates of the mean, variance and QoE respectively. Note that OPTAVR($\mathbf{m}, \mathbf{v}, c$) is closely related to OPT-ONLINE (discussed in Subsection I-A). Let $\mathbf{r}^*(\mathbf{m}, \mathbf{v}, c)$ denote the optimal solution to OPTAVR($\mathbf{m}, \mathbf{v}, c$).

Let

$$\mathcal{H} := [0, r_{\max}]^N \times [0, v_{\max}]^N, \quad (17)$$

where $\times$ denotes Cartesian product operator for sets. Next, we describe our algorithm in detail.

**Algorithm 1. Adaptive Variance aware Reward allocation (AVR)**

**AVR.0**: Initialization: let $(\mathbf{m}(1), \mathbf{v}(1)) \in \mathcal{H}$.

In each time slot $t \in \mathbb{N}$, carry out the following steps:
**AVR.1**: The reward allocation in time slot $t$ is the optimal solution to OPTAVR($\mathbf{m}(t), \mathbf{v}(t), c_t$), i.e., $\mathbf{r}^*(\mathbf{m}(t), \mathbf{v}(t), c_t)$, and will be denoted by $\mathbf{r}^*(t)$ (when the dependence on the variables is clear from context).
**AVR.2**: In time slot $t$, update $m_i$ as follows: for all $i \in \mathcal{N}$,

$$m_i(t+1) = \left[m_i(t) + \frac{1}{t}\left(r_i^*(t) - m_i(t)\right)\right]_0^{r_{\max}}, \quad (18)$$

and update $v_i$ as follows: for all $i \in \mathcal{N}$,

$$v_i(t+1) = \left[v_i(t) + \frac{(r_i^*(t) - m_i(t))^2 - v_i(t)}{t}\right]_0^{v_{\max}} \quad (19)$$

Here, $[x]_a^b = \min(\max(x,a), b)$.

We see that the update equations (18)-(19) roughly ensure that the parameters $\mathbf{m}(t)$ and $\mathbf{v}(t)$ keep track of mean reward and variance in reward allocation respectively associated with the reward allocation under AVR. Also, note that we do not have to keep track of the estimates of variance of users $i$ with linear $U_i^V$ since OPTAVR is insensitive to their values (i.e., $\left(U_i^V\right)'(.)$ is a constant), and thus the evolutions of $\mathbf{m}(t)$ and $(v_i(t))_{i \in \mathcal{N}_n}$ do not depend on them. We let $\boldsymbol{\theta}(t) = (\mathbf{m}(t), \mathbf{v}(t))$ for each $t$. The update equations (18)-(19) ensure that $\boldsymbol{\theta}(t)$ stays in the set $\mathcal{H}$.

For any $(\mathbf{m}, \mathbf{v}, c) \in \mathcal{H} \times \mathcal{C}$, we have $\left(U_i^E\right)'\left(m_i - U_i^V(v_i)\right)\left(U_i^V\right)'(v_i) > 0$ for each $i \in \mathcal{N}$

(see Assumptions U.E and U.V). Hence, OPTAVR($\mathbf{m}, \boldsymbol{v}, c$) is a convex optimization problem with a unique solution. Further, using (2) in Assumption C.3, we can show that it satisfies Slater's condition. Hence, the optimal solution $\mathbf{r}^*$ for OPTAVR($\mathbf{m}, \boldsymbol{v}, c$) satisfies KKT conditions given below.

---

**KKT-OPTAVR**($\mathbf{m}, \boldsymbol{v}, c$):
There exist nonnegative constants $\mu^*$ and $(\gamma_i^*)_{i \in \mathcal{N}}$ such that for all $i \in \mathcal{N}$

$$\left(U_i^E\right)'\left(m_i - U_i^V(v_i)\right)\left(1 - 2\left(U_i^V\right)'(v_i)(r_i^* - m_i)\right)$$
$$+\gamma_i^* - \mu^* c_i'(\mathbf{r}^*) = 0, \quad (20)$$
$$\mu^* c(\mathbf{r}^*) = 0, \quad (21)$$
$$\gamma_i^* r_i^* = 0. \quad (22)$$

---

In the next lemma, we establish continuity properties of $\mathbf{r}^*(\mathbf{m}, \boldsymbol{v}, c)$ (also denoted by $\mathbf{r}^*$ in the result) when viewed as a function of $(\mathbf{m}, \boldsymbol{v})$. In particular, the Lipschitz assumption on the derivatives of $(U_i^V)_{i \in \mathcal{N}}$ and $(U_i^E)_{i \in \mathcal{N}}$ help us conclude that the optimizer of OPTAVR($\boldsymbol{\theta}, c$) is Lipschitz continuous in $\boldsymbol{\theta}$. A proof is given in Appendix A.

**Lemma 2.** *For any $c \in \mathcal{C}$, and $\boldsymbol{\theta} = (\mathbf{m}, \boldsymbol{v}) \in \mathcal{H}$*
*(a) $\mathbf{r}^*(\boldsymbol{\theta}, c)$ is a Lipschitz continuous function of $\boldsymbol{\theta}$.*
*(b) $E\left[\mathbf{r}^*\left(\boldsymbol{\theta}, C^\pi\right)\right]$ is a Lipschitz continuous function of $\boldsymbol{\theta}$.*

The next theorem states our result related to the convergence of the mean, variance and QoE of the reward allocations under AVR. This result is proven in Section V. For brevity, we let $\mathbf{r}^*(t)$ denote $\mathbf{r}^*(\mathbf{m}(t), \boldsymbol{v}(t), c_t)$.

**Theorem 1.** *The evolution of parameters $\mathbf{m}(t)$ and $\mathbf{v}(t)$, and the allocation $(\mathbf{r}^*)_{1:T}$ associated with AVR satisfies the following property: For almost all sample paths, and for each $i \in \mathcal{N}$,*

$$(a) \quad \lim_{T \to \infty} m^T(r_i^*) = \lim_{t \to \infty} m_i(t),$$
$$(b) \quad \lim_{T \to \infty} \mathrm{Var}^T(r_i^*) = \lim_{t \to \infty} v_i(t),$$
$$(c) \quad \lim_{T \to \infty} e_i^T(r_i^*) = \lim_{t \to \infty} \left(m_i(t) - U_i^V(v_i(t))\right).$$

*Asymptotic Optimality of AVR*:

The next result establishes the asymptotic optimality of AVR, i.e., if we consider long periods of time $T$, the difference in performance of AVR and the optimal offline policy OPT($T$) becomes negligible.

**Theorem 2.** *The allocation $(\mathbf{r}^*)_{1:T}$ associated with AVR is feasible, i.e., it satisfies* (6) *and* (7). *Also, for almost all sample paths AVR is asymptotically optimal, i.e.,*

$$\lim_{T \to \infty} \left(\phi_T(\mathbf{r}^*) - \phi_T(\mathbf{r}^T)\right) = 0.$$

*Proof:* Since the allocation $(\mathbf{r}^*)_{1:T}$ associated with AVR satisfies (15) and (16) at each time slot, it also satisfies (6) and (7).

To show asymptotic optimality, consider any realization of $(c)_{1:T}$. Let $(\boldsymbol{\mu}^*)_{1:T}$ and $(\boldsymbol{\gamma}^*)_{1:T}$ be the sequences of nonnegative real numbers satisfying (20), (21) and (22) for this realization. From the nonnegativity of these numbers, and feasibility of $(\mathbf{r}^T)_{1:T}$, we have

$$\phi_T\left(\mathbf{r}^T\right) \leq \psi_T\left(\mathbf{r}^T\right), \quad (23)$$

where

$$\psi_T\left(\mathbf{r}^T\right) = \sum_{i \in \mathcal{N}} U_i^E\left(e_i^T(r_i^T)\right)$$
$$- \sum_{t=1}^{T} \frac{\mu^*(t)}{T} c_t(\mathbf{r}^T(t)) + \sum_{t=1}^{T} \sum_{i \in \mathcal{N}} \frac{\gamma_i^*(t)}{T} r_i^T(t).$$

The function $\psi_T$ is the Lagrangian associated with OPT($T$) but evaluated at the optimal Lagrange multipliers associated with the optimization problems (OPTAVR) involved in AVR, and hence the inequality. Since $\psi_T$ is a differentiable concave function, we have (see [6])

$$\psi_T\left(\mathbf{r}^T\right) \leq \psi_T\left(\mathbf{r}^*\right)$$
$$+ \left\langle \nabla \psi_T\left(\mathbf{r}^*\right), \left((\mathbf{r}^T)_{1:T} - (\mathbf{r}^*)_{1:T}\right)\right\rangle,$$

where $\langle \cdot, \cdot \rangle$ denotes the dot product. Hence, we have

$$\psi_T\left(\mathbf{r}^T\right) \leq \sum_{i \in \mathcal{N}} U_i^E\left(e_i^T(r_i^*)\right) - \sum_{t=1}^{T} \frac{\mu^*(t)}{T} c_t(\mathbf{r}^*(t))$$
$$+ \sum_{t=1}^{T} \sum_{i \in \mathcal{N}} \frac{\gamma_i^*(t)}{T} r_i^*(t) + \sum_{t=1}^{T} \sum_{i \in \mathcal{N}} \left(r_i^T(t) - r_i^*(t)\right)$$
$$\left(-\frac{\mu^*(t)}{T} c_{t,i}'(\mathbf{r}^*(t)) + \frac{\gamma_i^*(t)}{T} + \left(U_i^E\right)'\left(e_i^T(r_i^*)\right)\right.$$
$$\left.\left(\frac{1}{T} - \frac{2\left(U_i^V\right)'\left(\mathrm{Var}^T(r_i^*)\right)}{T}\left(r_i^*(t) - m^T(r_i^*)\right)\right)\right).$$

Using (23), and the fact that $(\boldsymbol{\mu}^*)_{1:T}$ and $(\boldsymbol{\gamma}^*)_{1:T}$ satisfy (20), (21) and (22), we have

$$\phi_T\left(\mathbf{r}^T\right) \leq \sum_{i \in \mathcal{N}} U_i^E\left(e_i^T(r_i^*)\right) + \sum_{t=1}^{T} \sum_{i \in \mathcal{N}} \frac{r_i^T(t) - r_i^*(t)}{T}$$
$$\left(\left(U_i^E\right)'\left(e_i^T(r_i^*)\right)\left(1 - 2\left(U_i^V\right)'\left(\mathrm{Var}^T(r_i^*)\right)\right.\right.$$
$$\left.\left(r_i^*(t) - m^T(r_i^*)\right)\right)$$
$$- \left(U_i^E\right)'\left(e_i(t-1)\right)\left(1 - 2\left(U_i^V\right)'(v_i(t-1))\right.$$
$$\left.\left.\left(r_i^*(t) - m_i(t-1)\right)\right)\right). \quad (24)$$

From Theorem 1 (a)-(c), and the continuity and boundedness of the functions involved, we can conclude that the expression appearing in the last four lines of the above inequality can be made as small as desired by choosing large enough $T$ and then choosing a large enough $t$. Also, $\left|r_i^T(t) - r_i^*(t)\right| \leq r_{\max}$ for each $i \in \mathcal{N}$. Hence, taking limits in (24),

$$\lim_{T \to \infty} \left(\phi_T(\mathbf{r}^*) - \phi_T(\mathbf{r}^T)\right) \geq 0. \quad (25)$$

holds for almost all sample paths. From the optimality of $(\mathbf{r}^T)_{1:T}$,

$$\phi_T(\mathbf{r}^T) \geq \phi_T(\mathbf{r}^*). \tag{26}$$

The result follows from the inequalities (25) and (26). ∎

## V. CONVERGENCE ANALYSIS

This section is devoted to the proof of Theorem 1 which captures the convergence of reward allocations under AVR. We start the section by studying another optimization problem OPTSTAT closely related to OPT($T$).

### A. A stationary version of OPT: OPTSTAT

The formulation OPT(T) involves time averages of various quantities associated with users' rewards. By contrast, the formulation of OPTSTAT is based on expected values of the corresponding quantities under the stationary distribution of $(C_t)_{t \in \mathbb{N}}$.

Recall that (see C.1) $(C_t)_{t \in \mathbb{N}}$ is a stationary ergodic process with marginal distribution $(\pi(c))_{c \in \mathcal{C}}$, i.e., for $c \in \mathcal{C}$, $\pi(c)$ is the probability of the event $C_t = c$. Since $\mathcal{C}$ is finite, we assume that $\pi(c) > 0$ for each $c \in \mathcal{C}$ without any loss of generality. Let $(\boldsymbol{\rho}_c)_{c \in \mathcal{C}}$ be a vector (of vectors) where $\boldsymbol{\rho}_c \in \mathbb{R}^N$ is the reward allocation to the users for constraint $c \in \mathcal{C}$. Now, let

$$\phi_\pi\left((\boldsymbol{\rho}_c)_{c \in \mathcal{C}}\right) = \sum_{i \in \mathcal{N}} U_i^E\left(E\left[\rho_{C^\pi,i}\right] - U_i^V\left(\text{Var}\left(\rho_{C^\pi,i}\right)\right)\right)$$

where

$$E[\rho_{C^\pi,i}] = \sum_{c \in \mathcal{C}} \pi(c) \rho_{c,i},$$
$$\text{Var}(\rho_{C^\pi,i}) = \sum_{c \in \mathcal{C}} \pi(c) (\rho_{c,i} - E[\rho_{C^\pi,i}])^2.$$

We define the 'stationary' optimization problem OPTSTAT as follows:

$$\max_{(\boldsymbol{\rho}_c)_{c \in \mathcal{C}}} \quad \phi_\pi\left((\boldsymbol{\rho}_c)_{c \in \mathcal{C}}\right),$$
$$\text{subject to} \quad c(\boldsymbol{\rho}_c) \leq 0, \quad \forall\, c \in \mathcal{C},$$
$$\rho_{c,i} \geq 0, \quad \forall\, i \in \mathcal{N}, \quad \forall\, c \in \mathcal{C}.$$

The next lemma gives a few useful properties of OPTSTAT.

**Lemma 3.** *(a) OPTSTAT is a convex optimization problem satisfying Slater's condition.*
*(b) OPTSTAT has a unique solution.*

*Proof:* The proof is similar to that of Lemma 1, and is easy to establish once we prove the convexity of the function $\text{Var}(\cdot)$. ∎

Using Lemma 3 (a), we can conclude that KKT conditions given below are necessary and sufficient for optimality for OPTSTAT. Let $(\boldsymbol{\rho}_c^\pi)_{c \in \mathcal{C}}$ denote the optimal solution.

**KKT-OPTSTAT**:

There exist constants $(\mu^\pi(c))_{c \in \mathcal{C}}$ and $(\gamma^\pi(c))_{c \in \mathcal{C}}$ are such that

$$\pi(c)\left(U_i^E\right)'\left(E\left[\rho_{C^\pi,i}^\pi\right] - U_i^V\left(\text{Var}\left(\rho_{C^\pi,i}^\pi\right)\right)\right)$$
$$\left(1 - 2\left(U_i^V\right)'\left(\text{Var}\left(\rho_{C^\pi,i}^\pi\right)\right)\left(\rho_{c,i}^\pi - E\left[\rho_{C^\pi,i}^\pi\right]\right)\right)$$
$$-\mu^\pi(c)\, c_i'(\boldsymbol{\rho}_c^\pi) + \gamma_i^\pi(c) = 0, \tag{27}$$
$$\mu^\pi(c)\, c(\boldsymbol{\rho}_c^\pi) = 0, \tag{28}$$
$$\gamma_i^\pi(c)\, \rho_{c,i}^\pi = 0, \tag{29}$$

where $c_i'$ denotes $\frac{\partial c}{\partial \rho_i}$.

In developing the above KKT conditions, we used the fact that for any $c_0 \in \mathcal{C}$, $i \in \mathcal{N}$,

$$\frac{\partial \text{Var}\left(\rho_{C^\pi,i}^\pi\right)}{\partial \rho_{c_0,i}} = 2\pi(c_0)\left(\rho_{c_0,i}^\pi - E\left[\rho_{C^\pi,i}^\pi\right]\right).$$

Next, we find relationships between the optimal solution $(\boldsymbol{\rho}_c^\pi)_{c \in \mathcal{C}}$ of OPTSTAT and OPTAVR. To that end, for each $i \in \mathcal{N}$, let

$$m_i^\pi := E\left[\rho_{C^\pi,i}^\pi\right], \tag{30}$$
$$v_i^\pi := \text{Var}^\pi\left(\rho_{C^\pi,i}^\pi\right), \tag{31}$$
$$e_i^\pi := m_i^\pi - U_i^V(v_i^\pi). \tag{32}$$

**Definition 2.** Let

$$\mathcal{H}^* = \{(\mathbf{m}, \boldsymbol{v}) \in \mathcal{H} : (\mathbf{m}, \boldsymbol{v}) \text{ satisfies } (33) - (34)\},$$

where the conditions (33)-(34) are given below:

$$E\left[r_i^*(\mathbf{m}, \boldsymbol{v}, C^\pi)\right] = m_i \quad \forall\, i \in \mathcal{N}, \tag{33}$$
$$\text{Var}\left(r_i^*(\mathbf{m}, \boldsymbol{v}, C^\pi)\right) = v_i \quad \forall\, i \in \mathcal{N}. \tag{34}$$

Recall that $\mathbf{r}^*(\mathbf{m}, \boldsymbol{v}, c)$ is the optimal solution to OPTAVR$(\mathbf{m}, \boldsymbol{v}, c)$ and $\mathcal{H}$ is defined in (17).

The next result gives properties relating $(\mathbf{m}^\pi, \boldsymbol{v}^\pi)$ to the set $\mathcal{H}^*$. A proof is given in Appendix B.

**Theorem 3.** $(\mathbf{m}^\pi, \boldsymbol{v}^\pi)$ *satisfies the following:*
*(a)* $\mathbf{r}^*(\mathbf{m}^\pi, \boldsymbol{v}^\pi, c) = \boldsymbol{\rho}_c^\pi$ *for each $c \in \mathcal{C}$, and*
*(b)* $\mathcal{H}^* = \{(\mathbf{m}^\pi, \boldsymbol{v}^\pi)\}$.

In the above discussion, we identified several interesting relationships between OPTSTAT and OPTAVR, and identified some properties of the vectors $\mathbf{m}^\pi$, $\boldsymbol{v}^\pi$ and $\mathbf{e}^\pi$. Next, we use these to study a differential equation that mimics the evolution of the parameters in AVR.

### B. Dynamics of OPTAVR

In this subsection, we focus on establishing convergence of the following differential equation

$$\frac{d\boldsymbol{\theta}(\tau)}{d\tau} = \bar{\mathbf{g}}(\boldsymbol{\theta}(\tau)) + \mathbf{z}(\boldsymbol{\theta}(\tau)), \tag{35}$$

for $\tau \geq 0$ with $\boldsymbol{\theta}(0) \in \mathcal{H}$ where $\bar{\mathbf{g}}(\boldsymbol{\theta})$ is a function taking values in $\mathbb{R}^{2N}$ defined as follows: for $\boldsymbol{\theta} = (\mathbf{m}, \boldsymbol{v}) \in \mathcal{H}$, let

$$(\bar{\mathbf{g}}(\boldsymbol{\theta}))_i := E\left[r_i^*(\boldsymbol{\theta}, C^\pi)\right] - m_i, \tag{36}$$
$$(\bar{\mathbf{g}}(\boldsymbol{\theta}))_{N+i} := E\left[(r_i^*(\boldsymbol{\theta}, C^\pi) - m_i)^2\right] - v_i. \tag{37}$$

In (35), $z(\boldsymbol{\theta}) \in -C_{\mathcal{H}}(\boldsymbol{\theta})$ is a projection term corresponding to the smallest vector that ensures that the solution remains in $\mathcal{H}$ (see Section 4.3 of [17]). The set $C_{\mathcal{H}}(\boldsymbol{\theta})$ contains only the zero element when $\boldsymbol{\theta}$ is in interior of $\mathcal{H}$, and for $\boldsymbol{\theta}$ on the boundary of the set $\mathcal{H}$, $C_{\mathcal{H}}(\boldsymbol{\theta})$ is the convex cone generated by the outer normals at $\boldsymbol{\theta}$ of the faces of $\mathcal{H}$ on which $\boldsymbol{\theta}$ lies. The motivation for studying the above differential equation should be partly clear by comparing the RHS of (35) (see (36)-(37)) with the update equations in (18)-(19) in AVR, and we can associate the term $z(\boldsymbol{\theta})$ with the constrained nature of those update equations. The following result shows that $z(\boldsymbol{\theta})$ appearing in (35) is innocuous in the sense that we can ignore it when we study the differential equation. The proof, given in Appendix C, shows the redundancy of the term $z(\boldsymbol{\theta})$ by arguing that the differential equation itself ensures that $\boldsymbol{\theta}(\tau)$ stays within $\mathcal{H}$.

**Lemma 4.** *For any $\boldsymbol{\theta} \in \mathcal{H}$, $z_j(\boldsymbol{\theta}) = 0$ for all $1 \leq j \leq 2N$.*

Note that (35) has a unique solution for a given initialization due to Lipschitz continuity results in Lemma 2.

**Definition 3.** We say that an allocation scheme $(\boldsymbol{\rho}_c)_{c \in \mathcal{C}}$ is *feasible* if for each $c \in \mathcal{C}$, $c(\boldsymbol{\rho}_c) \leq 0$ and $\rho_{c,i} \geq 0$ for each $i \in \mathcal{N}$. Also, let $\mathcal{R}_{\mathcal{C}} \subset \mathbb{R}^{N|\mathcal{C}|}$ denote the set of feasible allocations, i.e.,

$$\mathcal{R}_{\mathcal{C}} := \Pi_{c \in \mathcal{C}} \left\{ \boldsymbol{\rho}_c \in \mathbb{R}^N : c(\boldsymbol{\rho}_c) \leq 0, \ \rho_{c,i} \geq 0 \quad \forall \ i \in \mathcal{N} \right\},$$

which corresponds to the set of feasible stationary policies.

We define the set $\widetilde{\mathcal{H}} \subset \mathcal{H}$ as follows:

$$\widetilde{\mathcal{H}} := \left\{ (\mathbf{m}, \boldsymbol{v}) \in \mathcal{H} : \text{ there exists } (\boldsymbol{\rho}_c)_{c \in \mathcal{C}} \in \mathcal{R}_{\mathcal{C}} \text{ such that} \right.$$
$$\left. E[\rho_{C^\pi, i}] = m_i, \ \text{Var}(\rho_{C^\pi, i}) \leq v_i \leq r_{\max}^2 \quad \forall \ i \in \mathcal{N} \right\}.$$

We can roughly think of $\widetilde{\mathcal{H}}$ as the set of all 'achievable' mean variance pairs. Here, the restriction $v_i \leq r_{\max}^2$ for each $i$ ensures that $\widetilde{\mathcal{H}}$ is bounded. Further, for any $\boldsymbol{\theta} = (\mathbf{m}, \boldsymbol{v}) \in \widetilde{\mathcal{H}}$, let

$$\begin{aligned} \widetilde{\mathcal{R}}(\boldsymbol{\theta}) := & \left\{ (\boldsymbol{\rho}_c)_{c \in \mathcal{C}} \in \mathcal{R}_{\mathcal{C}} : E[\rho_{C^\pi, i}] = m_i, \right. \\ & \left. \text{Var}(\rho_{C^\pi, i}) \leq v_i \quad \forall \ i \in \mathcal{N} \right\}. \end{aligned}$$

We can view $\widetilde{\mathcal{R}}(\boldsymbol{\theta})$ as the set of all feasible reward allocations corresponding to an achievable $\boldsymbol{\theta} \in \widetilde{\mathcal{H}}$.

The following result characterizes several useful properties of the sets introduced above; a proof is given in Appendix D.

**Lemma 5.** *(a) $\widetilde{\mathcal{R}}(\boldsymbol{\theta})$ is a non-empty compact subset of $\mathbb{R}^{N|\mathcal{C}|}$ for any $\boldsymbol{\theta} = (\mathbf{m}, \boldsymbol{v}) \in \widetilde{\mathcal{H}}$.*
*(b) $\widetilde{\mathcal{H}}$ is a bounded, closed and convex set.*

The next result gives a set of sufficient conditions to establish asymptotic stability of a point with respect to an ordinary differential equation. This result is a generalization of Theorem 4 in [29]. A proof of the result is given in Appendix E.

**Lemma 6.** *Consider a differential equation*

$$\dot{x} = f(x), \qquad x \in \mathbb{R}^d, \tag{38}$$

*where $f$ is locally Lipschitz and all trajectories exist for $t \in [0, \infty)$. Suppose that some compact set $K \subset \mathbb{R}^d$ is asymptotically stable with respect to (38) and also suppose that there exists a continuously differentiable function $L \colon \mathbb{R}^d \to \mathbb{R}$ and some $x_0 \in K$ such that*

$$\nabla L(x) \cdot f(x) < 0 \qquad \forall x \in K, \ x \neq x_0. \tag{39}$$

*Then $x_0$ is an asymptotically stable equilibrium for (38) in $\mathbb{R}^d$.*

In the next result, we establish a convergence result for the ODE in (35). The proof relies on the optimality properties of the solutions to OPTAVR, Lemma 3 from [29] a Theorem 3 (b), and Lemma 6. A detailed proof is given in Appendix F.

**Theorem 4.** *Suppose $\boldsymbol{\theta}(\tau)$ evolves according to the ODE (35). Then, for any initial condition $\boldsymbol{\theta}(0) \in \mathcal{H}$, $\lim_{\tau \to \infty} \boldsymbol{\theta}(\tau) = \boldsymbol{\theta}^\pi$.*

### C. Convergence properties of AVR

In this subsection, we discuss the proof of Theorem 1. We first establish a convergence result for $(\boldsymbol{\theta}(t))_{t \in \mathbb{N}}$ using the convergence result for the differential equation (35). We do so by viewing (18)-(19) as a stochastic approximation update equation, using a result from [17] that helps us to relate it the ODE (35), and establishing the desired convergence result by utilizing the corresponding result obtained for the ODE in Theorem 4. A detailed proof of the result is given in Appendix G.

**Lemma 7.** *If $\boldsymbol{\theta}(0) \in \mathcal{H}$, then the sequence $(\boldsymbol{\theta}(t))_{t \in \mathbb{N}}$ generated by the Algorithm AVR converges almost surely to $\boldsymbol{\theta}^\pi$.*

Now we can prove Theorem 1 mainly using Lemma 7, and stationarity and ergodicity Assumptions. The detailed arguments are given in Appendix H

## VI. CONCLUSIONS

This work presents an important generalization of NUM framework to account for the deleterious impact of temporal variability allowing for tradeoffs between mean, fairness and variability associated with reward allocations across a set of users. We proposed a simple asymptotically optimal online algorithm AVR to solve problems falling in this framework. We believe such extensions to capture variability in resource allocations can be relevant to a fairly wide variety of systems.

## REFERENCES

[1] R. Agrawal and V. Subramanian. Optimality of certain channel aware scheduling policies. In *Proc. of 2002 Allerton Conference on Communication, Control and Computing*, 2002.
[2] D. Bertsekas. *Dynamic Programming and Optimal Control (2 Vol Set)*. Athena Scientific, 3rd edition, Jan 2007.
[3] F. Bonnans, J. F. Bonnans, and A. D. Ioffe. Quadratic growth and stability in convex programming problems with multiple solutions, 1995.
[4] J. F. Bonnans and A. Shapiro. *Perturbation analysis of optimization problems*. Springer, 2000.
[5] A. Bouch and M. A. Sasse. Network quality of service: What do users need? In *Proceedings of the 4th International Distributed Conference, 22 nd 23 rd*, pages 21–23, 1999.

## APPENDIX

### A. Proof of Lemma 2

*Proof:* For $\boldsymbol{\theta} = (\mathbf{m}, \mathbf{v})$, let

$$\Phi_{\boldsymbol{\theta}}(\mathbf{r}) := \sum_{i \in \mathcal{N}} \left(U_i^E\right)'(e_i)\left(r_i - \left(U_i^V\right)'(v_i)(r_i - m_i)^2\right) \quad (40)$$

for $\mathbf{r} \in \mathbb{R}^N$ where $e_i = m_i - U_i^V(v_i)$ for each $i \in \mathcal{N}$. Next, for any $\boldsymbol{\theta}^a, \boldsymbol{\theta}^b \in \mathcal{H}$ and $\mathbf{r} \in [-2r_{\max}, 2r_{\max}]^N$ (any optimal solution to OPTAVR, i.e., minimizer of $\Phi_{\boldsymbol{\theta}}(\mathbf{r})$ subject to constraints is an interior point of this set), let

$$\Delta\Phi(\mathbf{r}, \boldsymbol{\theta}^a, \boldsymbol{\theta}^b) = \Phi_{\boldsymbol{\theta}^b}(\mathbf{r}) - \Phi_{\boldsymbol{\theta}^a}(\mathbf{r}).$$

We prove part (a) (i.e., the Lipschitz continuity with respect to $\boldsymbol{\theta}$ of the optimizer $\mathbf{r}^*(\boldsymbol{\theta}, c)$ of $\Phi_{\boldsymbol{\theta}}(\mathbf{r})$ subject to constraint $c$) using Proposition 4.32 in [4]. The first condition in the Proposition requires that $\Delta\Phi(\cdot, \boldsymbol{\theta}^a, \boldsymbol{\theta}^b)$ be Lipschitz continuous. To show this, note that for any $\mathbf{r}^c, \mathbf{r}^d \in [-2r_{\max}, 2r_{\max}]^N$

$$\begin{aligned}
&\Delta\Phi(\mathbf{r}^c, \boldsymbol{\theta}^a, \boldsymbol{\theta}^b) - \Delta\Phi(\mathbf{r}^d, \boldsymbol{\theta}^a, \boldsymbol{\theta}^b) \\
&= \sum_{i \in \mathcal{N}} \left(\left(U_i^E\right)'(e_i^a) - \left(U_i^E\right)'(e_i^b)\right)(r_i^c - r_i^d) \\
&\quad + \sum_{i \in \mathcal{N}} \left(U_i^E\right)'(e_i^a)\left(U_i^V\right)'(v_i^a)(r_i^d - r_i^c)(r_i^d + r_i^c - 2m_i^a) \\
&\quad - \sum_{i \in \mathcal{N}} \left(U_i^E\right)'(e_i^b)\left(U_i^V\right)'(v_i^b)(r_i^d - r_i^c)(r_i^d + r_i^c - 2m_i^b).
\end{aligned}$$

Using the above expression, Lipschitz continuity and boundedness of $\left(U_i^{V'}\right)_{i \in \mathcal{N}}$ and $\left(U_i^{E'}\right)_{i \in \mathcal{N}}$ (see Assumptions U.V.1 and U.E), and boundedness of $\mathbf{r}^a$ and $\mathbf{r}^b$, we can conclude that there exists some positive finite constant $\eta$ such that

$$\Delta\Phi(\mathbf{r}^c, \boldsymbol{\theta}^a, \boldsymbol{\theta}^b) \leq \eta d(\boldsymbol{\theta}^a, \boldsymbol{\theta}^b) d(\mathbf{r}^a, \mathbf{r}^b).$$

Next, we establish the second condition given in the proposition referred to as second order growth condition. For this we use Theorem 6.1 (vi) from [3], and consider the functions $L$ and $\psi$ discussed in the exposition of the theorem. We have

$$L(\mathbf{r}, \boldsymbol{\theta}, \mu, \boldsymbol{\gamma}, c) = \Phi_{\boldsymbol{\theta}}(\mathbf{r}^*) - \Phi_{\boldsymbol{\theta}}(\mathbf{r}) + \mu c(\mathbf{r}) - \sum_{i \in \mathcal{N}} \gamma_i r_i,$$

and for $\mathbf{d} \in \mathbb{R}^{\mathcal{N}}$, we have

$$\psi_{\mathbf{r}^*(\boldsymbol{\theta}^a, c)}(\mathbf{d}) = \mathbf{d}^{tr} \nabla_{\mathbf{r}}^2 L(\mathbf{r}^*(\boldsymbol{\theta}^a, c), \boldsymbol{\theta}^a, \mu^m(c), \boldsymbol{\gamma}^m(c), c) \mathbf{d}$$

where $\mu^m(c)$ and $(\gamma_i^m(c) : i \in \mathcal{N})$ are Lagrange multipliers associated with the optimal solution to OPTAVR$(\boldsymbol{\theta}^a, c)$. Then, using convexity of $c$ we have

$$\psi_{\mathbf{r}^*(\boldsymbol{\theta}^a, c)}(\mathbf{d}) \geq \sum_{i \in \mathcal{N}} 2\left(U_i^E\right)'(e_i^a)\left(U_i^V\right)'(v_i^a) d_i^2.$$

Since $\left(U_i^{V'}\right)_{i \in \mathcal{N}}$ and $\left(U_i^{E'}\right)_{i \in \mathcal{N}}$ are strictly positive (see Assumptions U.V.1 and U.E), we can conclude that there exists some positive finite constant $\eta_1$ such that

$$\psi_{\mathbf{r}^*(\boldsymbol{\theta}^a, c)}(\mathbf{d}) \geq \eta_1 \|\mathbf{d}\|^2.$$

Now, using Theorem 6.1 (vi) from [3], we can conclude that second order growth condition is satisfied.

Thus, we have verified the conditions given in Proposition 4.32 in [4], and thus (a) holds. Then, (b) follows from (a) since $\mathcal{C}$ is finite and

$$E[\mathbf{r}^*(\boldsymbol{\theta}, C^\pi)] = \sum_{c \in \mathcal{C}} \pi(c)\mathbf{r}^*(\boldsymbol{\theta}, c).$$

∎

## B. Proof of Theorem 3

*Proof:* By KKT-OPTSTAT $(\boldsymbol{\rho}_c^\pi : c \in \mathcal{C})$, $(\mu^\pi(c) : c \in \mathcal{C})$ and $((\gamma_i^\pi(c))_{i \in \mathcal{N}} : c \in \mathcal{C})$ satisfy (27)-(29). To show that $\mathbf{r}^*(\mathbf{m}^\pi, \boldsymbol{v}^\pi, c) = \boldsymbol{\rho}_c^\pi$, we verify that $\boldsymbol{\rho}_c^\pi$ satisfies KKT-OPTAVR$(\mathbf{m}^\pi, \boldsymbol{v}^\pi, c)$. To that end, we can verify that $\boldsymbol{\rho}_c^\pi$ along with $\mu^* = \frac{\mu^\pi(c)}{\pi(c)}$ and $\left(\gamma_i^* = \frac{\gamma_i^\pi(c)}{\pi(c)} : i \in \mathcal{N}\right)$ satisfy (20)-(22) by using (27)-(29). This proves part (a).

To prove part (b), first note that $(\mathbf{m}^\pi, \boldsymbol{v}^\pi) \in \mathcal{H}^*$ and this follows from (a) and the definitions (see (30)-(31)) of $\mathbf{m}^\pi$ and $\boldsymbol{v}^\pi$. Next, note that for any $(\mathbf{m}, \boldsymbol{v}) \in \mathcal{H}^*$ and each $c \in \mathcal{C}$, $\mathbf{r}^*(\mathbf{m}, \boldsymbol{v}, c)$ is an optimal solution to OPTAVR and thus, there exist nonnegative constants $\mu^*(c)$ and $(\gamma_i^*(c) : i \in \mathcal{N})$ such that for all $i \in \mathcal{N}$, and satisfies KKT-OPTAVR given in (20)-(22). Also, since $(\mathbf{m}, \boldsymbol{v}) \in \mathcal{H}^*$, it satisfies (33)-(34). Combining these observations, we have that for all $c \in \mathcal{C}$

$$\begin{aligned}
\left(U_i^E\right)'\left(E\left[\mathbf{r}^*(\boldsymbol{\theta}, C^\pi)\right] - U_i^V\left(\text{Var}^\pi\left(\mathbf{r}^*(\boldsymbol{\theta}, C^\pi)\right)\right)\right)& \\
\left(r_i^*(\boldsymbol{\theta}, c) - 2\left(U_i^V\right)'\left(\text{Var}^\pi\left(\mathbf{r}^*(\boldsymbol{\theta}, C^\pi)\right)\right)\left(r_i^*(\boldsymbol{\theta}, c)\right.\right.& \\
\left.\left. - E\left[\mathbf{r}^*(\boldsymbol{\theta}, C^\pi)\right]\right]\right) + \gamma_i^* - \mu^*(c) c_i'(\mathbf{r}^*(\boldsymbol{\theta}, c)) &= 0, \\
\mu^*(c) c(\mathbf{r}^*(\boldsymbol{\theta}, c)) &= 0, \\
\gamma_i^* r_i^*(\boldsymbol{\theta}, c) &= 0.
\end{aligned}$$

where $\boldsymbol{\theta} = (\mathbf{m}, \boldsymbol{v})$, and $e_i = m_i - U_i^V(v_i)$ for each $i \in \mathcal{N}$. Now for each $c \in \mathcal{C}$, multiply the above equations with $\pi(c)$ and one obtains KKT-OPTSTAT ((27)-(29)) with $(\pi(c)\mu^*(c) : c \in \mathcal{C})$ and $((\pi(c)\gamma_i^*(c))_{i \in \mathcal{N}} : c \in \mathcal{C})$ as associated Lagrange multipliers. From Lemma 3 (a), OPTSTAT satisfies Slater's condition and hence satisfying KKT conditions is sufficient for optimality for OPTSTAT. Thus, we have that $(\mathbf{r}^*(\mathbf{m}, \boldsymbol{v}, c))_{c \in \mathcal{C}}$ is an optimal solution to OPTSTAT. This observation along with uniqueness of solution to OPTSTAT and (33)-(34), imply part (b), i.e., $\mathcal{H}^* = \{(\mathbf{m}^\pi, \boldsymbol{v}^\pi)\}$. ∎

## C. Proof of Lemma 4

*Proof:* Recall that $\mathcal{H} = [0, r_{\max}]^N \times [0, v_{\max}]^N$ and $v_{\max} = r_{\max}^2$. Note that for any $\boldsymbol{\theta}$ in the interior of $\mathcal{H}$, $z_j(\boldsymbol{\theta}) = 0$ for all $j$ such that $1 \leq j \leq 2N$ from the definition of $C_\mathcal{H}(\boldsymbol{\theta})$ and thus we can restrict our attention to the boundary of $\mathcal{H}$. For any $\boldsymbol{\theta}$ on the boundary of $\mathcal{H}$ and $i \in \mathcal{N}$, we can use the facts that $(\bar{\mathbf{g}}(\boldsymbol{\theta}))_i = E[r_i^*(\boldsymbol{\theta}, C^\pi)] - m_i$ and $0 \leq r_i^*(\boldsymbol{\theta}, C^\pi), m_i \leq r_{\max}$, to conclude that $z_i(\boldsymbol{\theta}) = 0$. Similarly, since $v_{\max} = r_{\max}^2$, we can show that $z_j(\boldsymbol{\theta}) = 0$ for any $j$ such that $N+1 \leq j \leq 2N$. ∎

## D. Proof of Lemma 5

*Proof:* For any $\boldsymbol{\theta} \in \widetilde{\mathcal{H}}$, using the definition of $\widetilde{\mathcal{H}}$, we see that $\widetilde{\mathcal{R}}(\boldsymbol{\theta})$ is a non-empty set. For any $c \in \mathcal{C}$, the set $\{\boldsymbol{\rho}_c \in \mathbb{R}^N : c(\boldsymbol{\rho}_c) \leq 0, \ \rho_{c,i} \geq 0 \ \forall \ i \in \mathcal{N}\}$ is compact due to continuity (see Assumption C.1) and boundedness (see Assumption C.2) of feasible region associated with functions in $\mathcal{C}$. Thus, $\mathcal{R}_\mathcal{C}$ is also compact. Now, note that $\widetilde{\mathcal{R}}(\boldsymbol{\theta})$ is the intersection of a compact set $\mathcal{R}_\mathcal{C}$, and Cartesian product of intersection of inverse images of closed sets associated with continuous functions (corresponding to $\mathbb{E}[.]$ and $\text{Var}(\cdot)$) defined over $\mathbb{R}^N$. Thus, $\widetilde{\mathcal{R}}(\boldsymbol{\theta})$ is compact, and this proves (a).

$\widetilde{\mathcal{H}}$ is bounded since $0 \leq m_i \leq r_{\max}$ and $0 \leq v_i \leq r_{\max}^2$ for each $i \in \mathcal{N}$, and each $(\mathbf{m}, \boldsymbol{v}) \in \widetilde{\mathcal{H}}$.

Let $(\overline{\mathbf{m}}, \overline{\boldsymbol{v}})$ be any limit point of $\widetilde{\mathcal{H}}$. Then, there exists a sequence $((\mathbf{m}_n, \boldsymbol{v}_n))_{n \in \mathbb{N}} \subset \widetilde{\mathcal{H}}$, such that $\lim_{n \to \infty} (\mathbf{m}_n, \boldsymbol{v}_n) = (\overline{\mathbf{m}}, \overline{\boldsymbol{v}})$. Let $(\boldsymbol{\rho}_{c,n})_{c \in \mathcal{C}} \in \widetilde{\mathcal{R}}((\mathbf{m}_n, \boldsymbol{v}_n))$ for each $n \in \mathbb{N}$. Since $((\boldsymbol{\rho}_{c,n})_{c \in \mathcal{C}})_{n \in \mathbb{N}}$ is a sequence in the compact set $\mathcal{R}_\mathcal{C}$, it has some convergent subsequence $((\boldsymbol{\rho}_{c,n_k})_{c \in \mathcal{C}})_{k \in \mathbb{N}}$. Suppose that the subsequence converges to $(\overline{\boldsymbol{\rho}}_c)_{c \in \mathcal{C}} \in \mathcal{R}_\mathcal{C}$. Then,

$$E\left[\overline{\rho}_{C^\pi, i}\right] = \lim_{k \to \infty} E\left[\rho_{C^\pi, n_k i}\right] = \lim_{k \to \infty} m_{n_k i} = \overline{m}_i,$$
$$\text{Var}\left(\overline{\rho}_{C^\pi, i}\right) = \lim_{k \to \infty} \text{Var}\left(\rho_{C^\pi, n_k i}\right) \leq \lim_{k \to \infty} v_{n_k i} = \overline{v}_i.$$

Thus, $(\overline{\boldsymbol{\rho}}_c)_{c \in \mathcal{C}} \in \widetilde{\mathcal{R}}((\overline{\mathbf{m}}, \overline{\boldsymbol{v}}))$, and hence, $(\overline{\mathbf{m}}, \overline{\boldsymbol{v}}) \in \widetilde{\mathcal{H}}$. Thus, $\widetilde{\mathcal{H}}$ contains all its limit points and hence is closed.

To show convexity, consider $(\mathbf{m}_1, \boldsymbol{v}_1), (\mathbf{m}_2, \boldsymbol{v}_2) \in \widetilde{\mathcal{H}}$, and we show that for any given $\alpha \in [0, 1]$, we have $\alpha(\mathbf{m}_1, \boldsymbol{v}_1) + (1 - \alpha)(\mathbf{m}_2, \boldsymbol{v}_2) \in \widetilde{\mathcal{H}}$. Let $(\boldsymbol{\rho}_{c,1})_{c \in \mathcal{C}} \in \widetilde{\mathcal{R}}((\mathbf{m}_1, \boldsymbol{v}_1))$ and $(\boldsymbol{\rho}_{c,2})_{c \in \mathcal{C}} \in \widetilde{\mathcal{R}}((\mathbf{m}_2, \boldsymbol{v}_2))$. Hence, $\text{Var}(r_{1i}(C^\pi)) \leq v_{1i}$, $\text{Var}(r_{2i}(C^\pi)) \leq v_{2i} \ \forall \ i \in \mathcal{N}$. Let $\boldsymbol{\rho}_{c,3} = \alpha \boldsymbol{\rho}_{c,1} + (1 - \alpha)\boldsymbol{\rho}_{c,2}$. Thus, for each $i \in \mathcal{N}$,

$$E\left[\rho_{C^\pi, 3i}\right] = \alpha m_1 + (1 - \alpha) m_2. \quad (41)$$

Next, note that $\text{Var}(\rho_{C^\pi})$ is a convex function of $(\boldsymbol{\rho}_c)_{c \in \mathcal{C}}$. This can be shown using convexity of square function and linearity of expectation. Thus, for each $i \in \mathcal{N}$,

$$\begin{aligned}
\text{Var}(\rho_{C^\pi, 3i}) &\leq \alpha \text{Var}(\rho_{C^\pi, 1i}) + (1 - \alpha) \text{Var}(\rho_{C^\pi, 2i}) \\
&\leq \alpha v_{1i} + (1 - \alpha) v_{2i}. \quad (42)
\end{aligned}$$

Thus, from (41) and (42), we have that $(\mathbf{r}_3(c))_{c \in \mathcal{C}} \in \widetilde{\mathcal{R}}(\alpha(\mathbf{m}_1, \boldsymbol{v}_1) + (1 - \alpha)(\mathbf{m}_2, \boldsymbol{v}_2))$, and thus $\alpha(\mathbf{m}_1, \boldsymbol{v}_1) + (1 - \alpha)(\mathbf{m}_2, \boldsymbol{v}_2) \in \widetilde{\mathcal{H}}$. ∎

## E. Proof of Lemma 6

*Proof:* The approach used here is similar to that in [29]. Let $\delta > 0$ be given. With $B_\delta(x_0)$ denoting the open ball of radius $\delta$ centered at $x_0$ select $\varepsilon \in (0, \delta)$ such that

$$\max_{\overline{B_\varepsilon}(x_0)} L < \min_{K \setminus B_\delta(x_0)} L. \quad (43)$$

This is possible, since the hypotheses imply that $L(x_0) < L(x)$ for all $x \in K$, $x \neq x_0$. Indeed, consider any solution $\gamma$ of (38) starting at $x \in K$, with $x \neq x_0$. Then the invariance of $K$ and (39) imply that the set of $\omega$-limit points of $\gamma$ is necessarily the singleton $\{x_0\}$. Note that $L$ is non-increasing along trajectories in $K$ and is strictly decreasing along any portion of a trajectory which does not contain $x_0$. Choose any $t' > 0$ such $\gamma(t) \neq x_0$ for all $t \in [0, t']$ (this is of course possibly by the continuity of $t \mapsto \gamma(t)$). Therefore we must have

$$L(x) = L(\gamma(0)) > L(\gamma(t')) \geq \lim_{t \to \infty} L(\gamma(t)) = L(x_0).$$

Since $K$ is asymptotically stable there exists a decreasing sequence of open sets $\{G_k\}_{k \in \mathbb{N}}$ such that each $G_k$ is invariant with respect to (38) and $\cap_{k \in \mathbb{N}} G_k = K$. By (39)–(43) and the continuity of $L$ and $\nabla L \cdot f$ we can select $n \in \mathbb{N}$ large enough such that

$$\nabla L(x) \cdot f(x) < 0 \qquad \forall x \in \bar{G}_n \setminus B_\varepsilon(x_0) \tag{44a}$$

$$\max_{\bar{B}_\varepsilon(x_0)} L < \min_{\bar{G}_n \setminus B_\delta(x_0)} L. \tag{44b}$$

It is clear by (44a)–(44b) that any trajectory starting in $G_n \cap B_\varepsilon(x_0)$ stays in $B_\delta(x_0)$, implying that $x_0$ is a stable equilibrium. Let $\gamma$ be any trajectory of (38). Asymptotic stability of $K$ implies that there exists $t_1 > 0$ such that $\gamma(t) \in G_n$ for all $t > t_1$. Also by (44a) there exists $t_2 \geq t_1$ such that $\gamma(t_2) \in G_n \cap B_\delta(x_0)$. Therefore $x_0$ is asymptotically stable. ∎

### F. Proof of Theorem 4

*Proof:* Applying Lemma 3 in [29] and by identifying $\mathcal{V} \equiv \tilde{\mathcal{H}}$, it follows that $\tilde{\mathcal{H}}$ is asymptotically stable for (32). Define

$$L(\boldsymbol{\theta}) = L(\mathbf{m}, \mathbf{v}) := -\sum_{i \in \mathcal{N}} U_i^E(m_i - U_i^V(v_i)).$$

Then

$$\nabla L(\boldsymbol{\theta}) \cdot \bar{g}(\boldsymbol{\theta}) = -\sum_{i \in \mathcal{N}} \left(U_i^E\right)'(m_i - U_i^V(v_i))\left(\mathbb{E}[r_i^*(\boldsymbol{\theta}, C^\pi)]\right)$$

$$-m_i - \left(U_i^V\right)'(v_i)\left(\mathbb{E}[(r_i^*(\boldsymbol{\theta}, C^\pi) - m_i)^2] - v_i\right)\right). \tag{45}$$

If $\boldsymbol{\theta} \in \tilde{\mathcal{H}}$, then for some $\boldsymbol{\rho} \in \tilde{\mathcal{R}}(\boldsymbol{\theta})$, (45) takes the form

$$\nabla L(\boldsymbol{\theta}) \cdot \bar{g}(\boldsymbol{\theta}) = -\mathbb{E}[\Phi_{\boldsymbol{\theta}}(\boldsymbol{r}^*(\boldsymbol{\theta}, C^\pi)) - \Phi_{\boldsymbol{\theta}}(\boldsymbol{\rho}_{C^\pi})] \tag{46}$$
$$-\sum_{i \in \mathcal{N}}\left(U_i^E\right)'(m_i - U_i^V(v_i))\left(U_i^V\right)'(v_i)(v_i - \mathrm{Var}(\rho_{C^\pi, i}))$$

where $\Phi_{\boldsymbol{\theta}}$ is defined in (40). The optimality of $r_i^*(\boldsymbol{\theta}, c)$ for OPTAVR$(\mathbf{m}, \mathbf{v}, c)$ and the fact that $\boldsymbol{\rho} \in \tilde{\mathcal{R}}(\boldsymbol{\theta})$ together with Assumptions U.V.1 and U.E. then imply that both terms on the right-hand-side of (46) are nonpositive and that they vanish only if

$$\mathbb{E}[r_i^*(\boldsymbol{\theta}, C^\pi)] = \mathbb{E}[\rho_{C^\pi, i}] = m_i, \tag{47}$$
$$\mathrm{Var}(r_i^*(\boldsymbol{\theta}, C^\pi)) = \mathrm{Var}(\rho_{C^\pi, i}) = v_i. \tag{48}$$

In turn, by Theorem 3 these imply that $\boldsymbol{\theta} = \boldsymbol{\theta}^\pi$. Therefore $\nabla L(\boldsymbol{\theta}) \cdot \bar{g}(\boldsymbol{\theta}) < 0$ for all $\boldsymbol{\theta} \in \tilde{\mathcal{H}}$, $\boldsymbol{\theta} \neq \boldsymbol{\theta}^\pi$ and the result follows by Lemmas 4 and 6. ∎

### G. Proof of Lemma 7

*Proof:* We start proving the result by viewing (18)-(19) as a stochastic approximation update equation, and using Theorem 1.1 of Chapter 6 from [17] to relate (18)-(19) to the ODE (35).

In the following, we show that all the Assumptions required to use the theorem are satisfied. The following sets, variables and functions $H$, $\boldsymbol{\theta}_t$, $\xi_t$, $\mathbf{Y}_t$, $\epsilon_t$, sigma algebras $\mathcal{F}_t$, $\boldsymbol{\beta}_t$, $\delta \mathbf{M}_t$ and the function $\mathbf{g}$ appearing in the exposition of Theorem 1.1 of [17], correspond to the following variables and functions in our problem setting: $H = \mathcal{H}$, $\boldsymbol{\theta}_t = (\mathbf{m}(t), \mathbf{v}(t))$, $\xi_t = c_t$, for each $i \in \mathcal{N}$ $(Y_t)_i = r_i^*(t) - m_i(t)$ and $(Y_t)_{i+N} = (r_i^*(t) - m_i(t))^2 - v_i(t)$, $\epsilon_t = \frac{1}{t}$ for each $t$, $\mathcal{F}_t$ is such that $(\boldsymbol{\theta}_0, \mathbf{Y}_{i-1}, \xi_i, i \leq t)$ is $\mathcal{F}_t$-measurable, $\boldsymbol{\beta}_t = \mathbf{0}$ and $\delta \mathbf{M}_t = \mathbf{0}$ for each $t$, $(g((\mathbf{m}, \mathbf{v}), c))_i = r_i^*(\mathbf{m}, \mathbf{v}, c) - m_i$ and $(g((\mathbf{m}, \mathbf{v}), c))_{i+N} = (r_i^*(\mathbf{m}, \mathbf{v}, c) - m_i)^2 - v_i$,

Equation (5.1.1) in [17] is satisfied due to our choice of $\epsilon_t$, and (A4.3.1) is satisfied due to our choice of $\mathcal{H}$. Further, (A.1.1) is satisfied as the solutions to OPTAVR are bounded. (A.1.2) holds due to the continuity result in Lemma 2 (a).

We next show that (A.1.3) holds by choosing the function $\bar{\mathbf{g}}$ as follows for each $i \in \mathcal{N}$: $(\bar{g}(\mathbf{m}, \mathbf{v}))_i = E[r_i^*(\mathbf{m}, \mathbf{v}, C^\pi)] - m_i$, and $(\bar{g}(\mathbf{m}, \mathbf{v}))_{i+N} = E\left[(r_i^*(\mathbf{m}, \mathbf{v}, C^\pi) - m_i)^2\right] - v_i$. Note that the continuity of the function $\bar{\mathbf{g}}$ follows from Lemma 2 (b).

From Section 6.2 of [17], if $\epsilon_t$ does not go to zero faster than the order of $\frac{1}{\sqrt{t}}$, for (A.1.3) to hold, we only need to show that the strong law of large numbers holds for $(g(\mathbf{m}, \mathbf{v}, C_t))_t$ for any $\hat{\mathbf{q}}$. The strong law of large numbers holds since $(C_t)_{t \in \mathbb{N}}$ is a stationary ergodic random process and $g$ is a bounded function. Assumptions (A.1.4) and (A.1.5) hold since $\boldsymbol{\beta}_t = \mathbf{0}$ and $\delta \mathbf{M}_t = \mathbf{0}$ for each $t$. To check (A.1.6) and (A.1.7), we use sufficient conditions discussed in [17] following Theorem 1.1. Assumption (A.1.6) holds since $g$ is bounded. (A.1.7) holds due to the continuity of $g((\mathbf{m}, \mathbf{v}), c)$ in $(\mathbf{m}, \mathbf{v})$ uniformly in $c$ which follows from the continuity result in Lemma 2 (a), and the finiteness of $\mathcal{C}$. Thus, using Theorem 1.1, we can conclude that on almost all sample paths, $(\boldsymbol{\theta}(t))_{t \in \mathbb{N}}$ converges to some limit set of the ODE (35) in $\mathcal{H}$. From Theorem 4, for any initialization in $\mathcal{H}$, this limit set is the singleton $\{\boldsymbol{\theta}^\pi\}$, and thus the main result follows. ∎

### H. Proving Theorem 1 using Lemma 7

For each $i \in \mathcal{N}$,

$$\frac{1}{T}\sum_{t=1}^T r_i^*(\boldsymbol{\theta}(t), C_t) = \frac{1}{T}\sum_{t=1}^T \left(r_i^*(\boldsymbol{\theta}(t), C_t) - r_i^*(\boldsymbol{\theta}^\pi, C_t)\right)$$
$$+ \frac{1}{T}\sum_{t=1}^T r_i^*(\boldsymbol{\theta}^\pi, C_t). \tag{49}$$

The first term of (49) converges to 0 a.s. (i.e., for almost all sample paths) as $T \to \infty$ by Lemma 7, the continuity of $\mathbf{r}^*(\boldsymbol{\theta}, c)$ in $\boldsymbol{\theta}$ (see Lemma 2 (a)) and the Dominated Convergence Theorem (see, for e.g., [11]). The second term converges to $\mathbb{E}[r_i^*(\boldsymbol{\theta}^\pi, C^\pi)]$ by Birkhoff's Ergodic Theorem (see, for e.g., [10]). Now, note that $\mathbb{E}[r_i^*(\boldsymbol{\theta}^\pi, C^\pi)] = m_i^\pi$ (see Theorem 3 (b) and (33)). Since by Lemma 7, $\lim_{t \to \infty} m_i(t) = m_i^\pi$, part (a) is proved.

Next, we prove part (b). Note that for each $i \in \mathcal{N}$,

$$\mathrm{Var}^T(r_i^*) = \frac{1}{T}\sum_{t=1}^T \left(r_i^*(\boldsymbol{\theta}(t), C_t) - \frac{1}{T}\sum_{s=1}^T r_i^*(\boldsymbol{\theta}(s), C_s)\right)^2$$
$$= \frac{1}{T}\sum_{t=1}^T \left(r_i^*(\boldsymbol{\theta}(t), C_t) - m_i^\pi\right)^2$$
$$- \left(\frac{1}{T}\sum_{s=1}^T r_i^*(\boldsymbol{\theta}(s), C_s) - m_i^\pi\right)^2. \tag{50}$$

The second term on the right-hand-side of (50) converges a.s. to zero as $t \to \infty$ by part (a). Also, following the same steps as in the proof of part (a), we see that the first term converges a.s. to $v_i^\pi$ as $T \to \infty$. Since by Lemma 7, $\lim_{t\to\infty} v_i(t) = v_i^\pi$, part (b) is proved.

Part (c) follows from parts (a) and (b).